\documentclass[letterpaper]{article} % DO NOT CHANGE THIS
\usepackage{aaai2026}  % DO NOT CHANGE THIS
\usepackage{times}  % DO NOT CHANGE THIS
\usepackage{helvet}  % DO NOT CHANGE THIS
\usepackage{courier}  % DO NOT CHANGE THIS
\usepackage[hyphens]{url}  % DO NOT CHANGE THIS
\usepackage{graphicx} % DO NOT CHANGE THIS
\usepackage{natbib}  % DO NOT CHANGE THIS AND DO NOT ADD ANY OPTIONS TO IT
\usepackage{caption} % DO NOT CHANGE THIS AND DO NOT ADD ANY OPTIONS TO IT
\usepackage{amsmath}
\usepackage{amssymb}
\usepackage{amsfonts}
\usepackage{algorithm}
\usepackage{algorithmic}
\usepackage{booktabs}
\usepackage{multirow}
\usepackage{pifont}
\usepackage{newfloat}
\usepackage{listings}
\DeclareCaptionStyle{ruled}{labelfont=normalfont,labelsep=colon,strut=off} % DO NOT CHANGE THIS
\floatstyle{ruled}
\newfloat{listing}{tb}{lst}{}
\floatname{listing}{Listing}
\title{Unsupervised Multi-Parameter Inverse Solving for Reducing Ring Artifacts in 3D X-Ray CBCT}
\author{
    Qing Wu\textsuperscript{\rm 1},
    Hongjiang Wei\textsuperscript{\rm 2},
    Jingyi Yu\textsuperscript{\rm 1},
    Yuyao Zhang\textsuperscript{\rm 1}\thanks{Corresponding author.}
}
\affiliations{
    \textsuperscript{\rm 1}School of Information Science and Technology, ShanghaiTech University, Shanghai 201210, China\\
    \textsuperscript{\rm 2}School of Biomedical Engineering, Shanghai Jiao Tong University, Shanghai 200127, China\\
    wuqing@shanghaitech.edu.cn, hongjiang.wei@sjtu.edu.cn, yujingyi@shanghaitech.edu.cn, zhangyy8@shanghaitech.edu.cn
}

\newcommand{\ie}{\textit{i}.\textit{e}.}
\newcommand{\eg}{\textit{e}.\textit{g}.}
\usepackage{bibentry}
\begin{document}

\maketitle

% \twocolumn[{
% \renewcommand\twocolumn[1][]{#1}
% \maketitle
% \begin{center}
% \captionsetup{type=figure}
% \includegraphics[width=0.7\textwidth]{fig/fig1_real_3d.pdf}
% \caption{We propose Riner, an unsupervised method for reducing ring artifacts in 3D X-ray CBCT imaging. On a real-world \textit{Chicken foot} sample (Fig.~\ref{fig:fig_real_device} \textit{Right}) with image dimension of 512$\times$512$\times$80 and an ultra-high resolution of 60$\times$60$\times$60 $\mu$m\textsuperscript{3}, acquired by a commercial Bruker SKYSCAN 1276 micro-CT scanner, our Riner effectively removes ring artifacts and reconstructs high-quality CT volumes, outperforming model-based algorithms (FDK~\cite{fdk} and Super~\cite{vo2018Super}) and SOTA supervised deep learning models (DeepRAR~\cite{trapp2022deeprar} and Restormer~\cite{zamir2022restormer})}
% \label{fig:3d_real}
% \end{center}
% }]

\begin{abstract}
Ring artifacts are prevalent in 3D cone-beam computed tomography (CBCT) due to non-ideal responses of X-ray detectors, substantially affecting image quality and diagnostic reliability. Existing state-of-the-art (SOTA) ring artifact reduction (RAR) methods rely on supervised learning with large-scale paired CT datasets. While effective in-domain, supervised methods tend to struggle to fully capture the physical characteristics of ring artifacts, leading to pronounced performance drops in complex real-world acquisitions. Moreover, their scalability to 3D CBCT is limited by high memory demands. In this work, we propose Riner, a new unsupervised RAR method. Based on a theoretical analysis of ring artifact formation, we reformulate RAR as a multi-parameter inverse problem, where the non-ideal responses of X-ray detectors are parameterized as solvable physical variables. Using a new differentiable forward model, Riner can jointly learn the implicit neural representation of artifact-free images and estimate the physical parameters directly from CT measurements, without external training data. Additionally, Riner is memory-friendly due to its ray-based optimization, enhancing its usability in large-scale 3D CBCT. Experiments on both simulated and real-world datasets show Riner outperforms existing SOTA supervised methods. 
\end{abstract}

\begin{links}
    \link{Code}{https://github.com/iwuqing/Riner}
    % \link{Datasets}{https://aaai.org/example/datasets}
    % \link{Extended version}{https://aaai.org/example/extended-version}
\end{links}
\section{Introduction}
\label{sec:intoduction}
\par
\begin{figure}[t]
    \centering
    \includegraphics[width=0.37\textwidth]{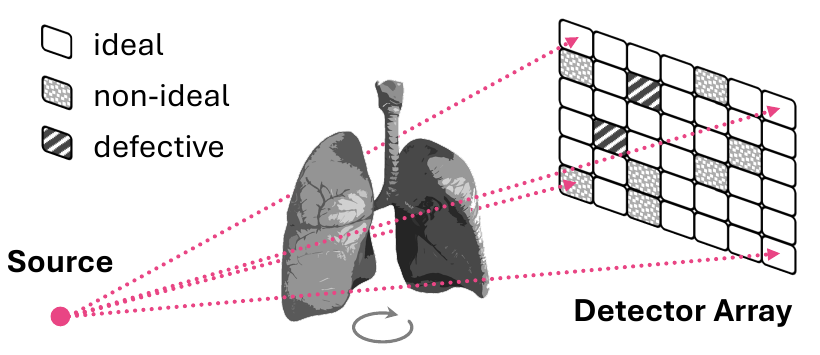}
    \caption{Illustration of 3D CBCT acquisition. An X-ray source emits cone-shaped beams that pass through objects and are received by a 2D detector array. Due to hardware limitations~\cite{boas2012ct}, the X-ray detectors can be ideal (accurate response), non-ideal (fluctuating response), or defective (no response).}
    \label{fig:fig_cbct}
\end{figure}

 3D cone-beam computed tomography (CBCT) enables detailed visualization of internal structures, providing critical information for applications ranging from medical diagnosis and biological research to materials science~\cite{scarfe2008cone, venkatesh2017cone}. CBCT acquisition involves a series of 2D projections captured by a 2D X-ray detector array at varying angles, as illustrated in Fig.~\ref {fig:fig_cbct}. However, due to inherent physical hardware limitations~\cite{boas2012ct}, individual detectors within the 2D array often struggle to maintain ideal, consistent signal responses, especially in $\mu$m-scale imaging using micro-CT scanners~\cite{yousuf2010efficient, rashid2012improved}. These non-ideal responses lead to nonlinear physical distortions in CT measurement data, suffering from severe ring artifacts when traditional linear reconstruction algorithms like FDK~\cite{fdk}, are applied. Such ring artifacts can markedly degrade the quality and reliability of resulting CT images.
\par Current supervised deep learning (DL) methods~\cite{hybridcnn, trapp2022deeprar, wang2023indudonet+, wang2022dudotrans} represent the state-of-the-art (SOTA) in the CT reconstructions. These methods train neural networks, such as UNet~\cite{ronneberger2015u} and ViT~\cite{dosovitskiy2020image}, on large-scale paired datasets to map artifact-corrupted images to artifact-free outputs. While promising, they often face two key limitations: \textit{1) High data collection costs and limited generalization}. Collecting paired CT images is often prohibitively expensive, especially for medical and biological applications. As a result, most models are trained on simulated datasets generated from clean CT images~\cite{fang2020removing, trapp2022deeprar}. However, simulation errors (\eg, differences in image resolution, detector characteristics, and CT geometry) can substantially degrade performance on unseen real-world data;  \textit{2) Scalability issues in 3D CBCT.} Supervised DL models typically are difficult to scale to 3D due to high memory requirements. Adapting 2D models by processing slice-by-slice often causes discontinuities along the Z-axis, compromising 3D volumetric consistency. These limitations confine the use of supervised DL methods to constrained research settings, limiting their broader applicability.
\begin{figure*}[t]
    \centering
    \includegraphics[width=0.8\textwidth]{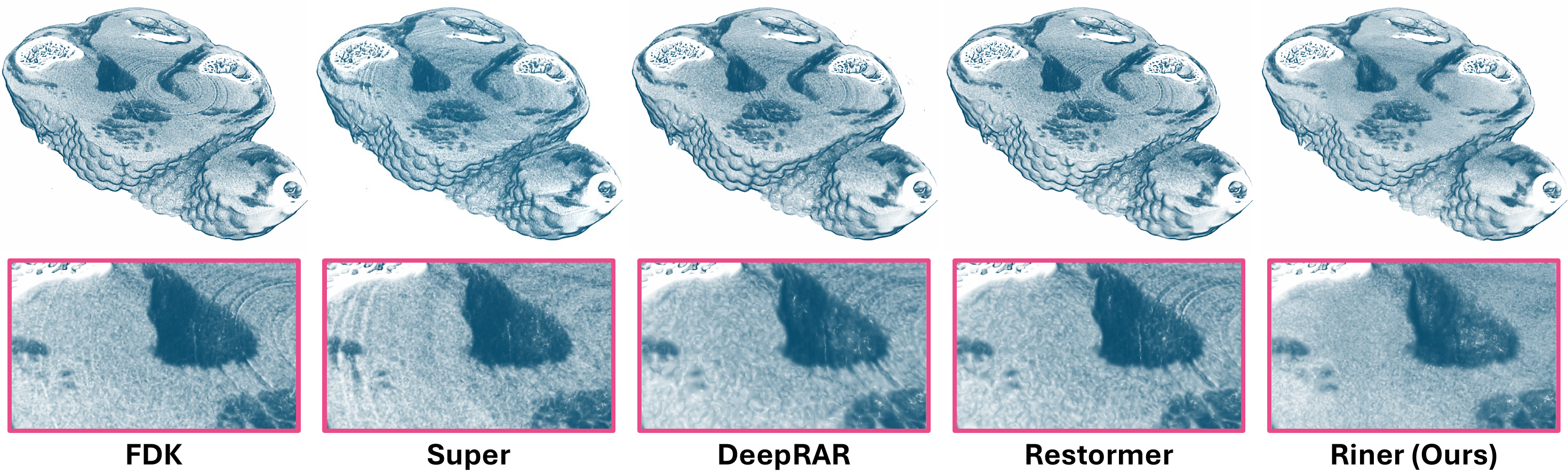}
    \caption{We propose Riner, an unsupervised method for reducing ring artifacts in 3D X-ray CBCT imaging. On a real-world \textit{Chicken foot} sample (see Appendix) with image dimension of 512$\times$512$\times$80 and an ultra-high resolution of 60$\times$60$\times$60 $\mu$m\textsuperscript{3}, acquired by a commercial Bruker SKYSCAN 1276 micro-CT scanner, our Riner effectively removes ring artifacts and reconstructs high-quality CT volumes, outperforming SOTA model-based algorithms (FDK~\cite{fdk} and Super~\cite{vo2018Super}) and SOTA supervised deep learning models (DeepRAR~\cite{trapp2022deeprar} and Restormer~\cite{zamir2022restormer})}
    \label{fig:3d_real}
\end{figure*}
\par In this paper, we propose Riner, a new unsupervised CT ring artifact reduction (RAR) method. Unlike end-to-end supervised learning, Riner reformulates RAR as a multi-parameter inverse problem grounded in X-ray CT physics. Theoretically, we analyze the physical origins of ring artifacts and identify two key nonlinear physical effects: inconsistent responses and invalid measurements. Motivated by the theoretical results, we introduce solvable physical parameters to characterize these two nonlinear effects. By leveraging a new differentiable forward model, Riner jointly learns the implicit neural representation (INR) of high-quality images and estimates the physical parameters directly from raw CT measurements. Estimating these parameters effectively corrects the two physical effects, improving image quality. While the inherent spectral bias of INR~\cite{rahaman2019spectral} helps regularize the ill-posed inverse problem, enabling high-quality CT reconstructions. Moreover, Riner uses a ray-based optimization scheme, making it memory-efficient and thus enhancing its usability in large-scale 3D CBCT settings.
\par We evaluate the proposed Riner on five datasets, including three simulated and two real-world cases, under both 2D fan-beam (FB) CT and 3D CBCT protocols. Experimental results show that our unsupervised method consistently outperforms existing SOTA techniques across all settings. Extensive ablation studies further validate the effectiveness of each key component in Riner. \textit{To the best of our knowledge, Riner is the first unsupervised RAR method to outperform supervised approaches.} 
\par The main contributions of this paper are as follows: 
\begin{itemize}
    \item We provide a theoretical analysis of the physical origins of CT ring artifacts and propose a novel multi-parameter solving framework to address the RAR problem.
    \item We propose Riner, a new unsupervised RAR method that jointly recovers high-quality CT images and estimates detector responses directly from raw measurements.
    \item We perform extensive experiments on both simulated and real-world data, confirming the superiority of our unsupervised Riner over SOTA RAR techniques.
\end{itemize}
\section{Related Work}
% \par In this section, we briefly review advances for ring artifact reduction in CT, multi-parameter inverse problem, and implicit neural representation for CT imaging.
\paragraph{Advances for Ring Artifact Reduction}
\label{sec:advances_for_ring_artifacts_removal}
\par Ring artifact reduction (RAR) remains a long-standing and challenging problem in X-ray CT imaging~\cite{chen2024research}. Traditional hardware-based correction methods~\cite{zhu2013micro} can deliver satisfactory performance, but they typically require specialized hardware designs and significantly increase system costs. To overcome these limitations, many studies have explored software-based solutions. Early model-based approaches~\cite{rivers1998Norm, wavefft, vo2018Super} rely on handcrafted filters to correct CT measurements, but their performance is often limited due to insufficient priors. More recently, supervised deep learning (DL) methods~\cite{hybridcnn, fang2020removing, trapp2022deeprar, wang2023indudonet+, wang2022dudotrans, zamir2022restormer, chen2022simple, guo2024mambair,guo2024mambairv2} have achieved SOTA results by learning data-driven priors from large-scale paired datasets. However, these end-to-end models often struggle to fully capture the underlying physical properties of ring artifacts, leading to poor generalization on out-of-domain (OOD) data. In contrast, we reformulate RAR as a multi-parameter inverse problem within a physics-driven framework to improve robustness and generalization.
\paragraph{Multi-Parameter Inverse Problem}
\label{sec:multi-parameter}
\par Multi-parameter inverse problems (MPIP) involve estimating multiple unknown parameters of a complex system from measurement data. By incorporating domain-specific physical priors, MPIP-based approaches have shown strong potential in a range of applications, including physics~\cite{dalla2022multi,cristofol2011uniqueness,zhu2023research}, partial differential equations~\cite{chen1984iterative}, and remote sensing~\cite{kostsov2015general}. Motivated by the success of this framework, we conduct a theoretical analysis of the physical mechanisms underlying CT ring artifact formation and reformulate the RAR problem as a multi-parameter inverse problem. This leads to Riner, a new unsupervised approach that improves CT reconstruction quality without relying on external training data.
\paragraph{Implicit Neural Representation for CT Imaging}
\label{sec:neural_field_for_medical_imaging}
\par Implicit neural representation (INR) is an unsupervised unified framework for solving visual inverse problems~\cite{essakine2024we}, where a multi-layer perceptron (MLP) is optimized to represent continuous signals of interest. Neural radiance fields~\cite{mildenhall2021nerf} primarily integrate volume rendering techniques into INR, achieving breakthrough progress in novel view synthesis. By incorporating a differentiable Radon transform to simulate the CT acquisition process, INR has been successfully extended to various CT tasks, such as sparse-view reconstruction~\cite{shen2022nerp, wu2023self, zha2022naf, ruckert2022neat, du2024dper, zang2021intratomo}, dynamic CT~\cite{reed2021dynamic, zhang2023dynamic, birklein2023neural}, and metal artifact reduction~\cite{wu2023unsupervised, wu2024unsupervised, lee2024neural}. However, existing INR-based methods do not address the challenging RAR problem, as they typically assume ideal X-ray detectors and thus ignore the nonlinear physical distortions caused by hardware imperfections. In contrast, our method introduces a new physical formulation that explicitly models non-ideal detector behavior by estimating additional parameters, enabling effective RAR reconstructions.
\begin{figure*}
    \centering
    \includegraphics[width=0.85\linewidth]{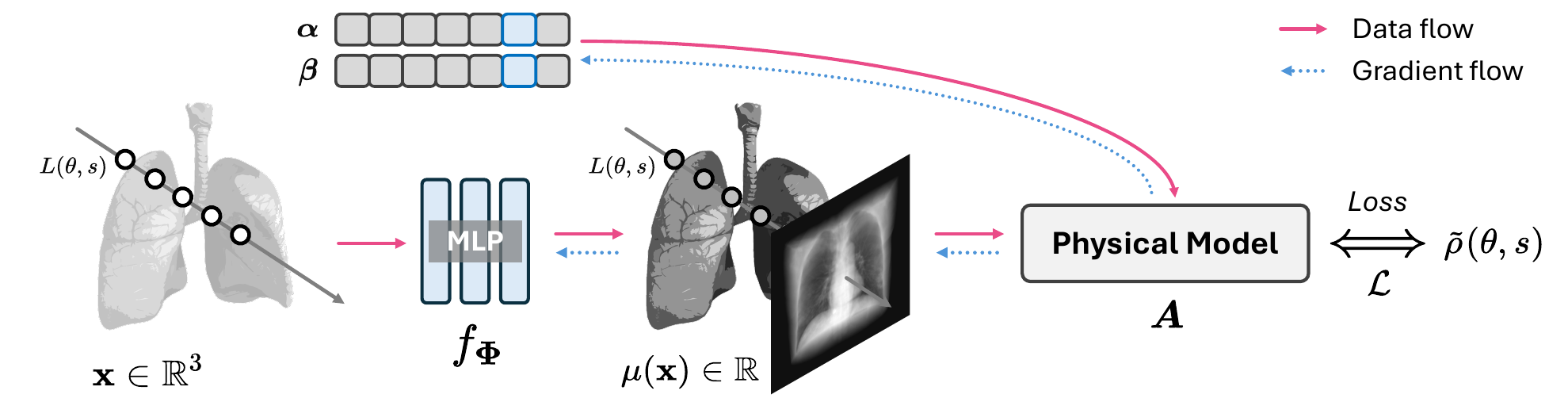}
    \caption{Overview of the proposed Riner method. Given raw measurements $\widetilde{\rho}(\theta, s)$, an MLP network $f_\mathbf{\Phi}$ receives multiple spatial coordinates $\mathbf{x}$ along an X-ray path $L(\theta,s)$ as input and predicts the corresponding CT intensities $\mu(\mathbf{x})=f_{\mathbf{\Phi}}(\mathbf{x})$. These predicted intensities, $\mu(\mathbf{x}),\forall\mathbf{x}\in L(\theta,s)$, the response factor $\alpha_s$, and the mask $\beta_s$ are then used to generate estimated measurements $\widehat{\rho}(\theta, s)$ via a differentiable physical model $\boldsymbol{A}$ (Eq.~\ref{eq:physical_model}). Finally, the MLP network $f_\mathbf{\Phi}$ and parameters $\alpha_s$, $\beta_s$ jointly are optimized by minimizing the loss function $\mathcal{L}$ (Eq.~\ref{eq:loss}) \textit{without} using any external data.}
    \label{fig:fig_method}
\end{figure*}
\section{Proposed Method}
% In this section, we first analyze the physical nature of CT ring artifacts, providing key insights for addressing RAR (Sec.~\ref{sec:theoritical_analyze}). Then, we present our problem formulation (Sec.~\ref{sec:problem_formulation}) and a new physical model (Sec.~\ref{sec:phyiscal_forward_model}). Finally, we introduce model optimization (Sec.~\ref{sec:model_optimization}), as shown in Fig.~\ref{fig:fig_method}.
\subsection{Theoretical Analysis for CT Ring Artifacts}
\label{sec:theoritical_analyze}
\par Lambert-Beer's law~\cite{lambert1760photometria,beer1852bestimmung} formulates the physical process of X-ray CT acquisition as below:
\begin{equation}
    I(\theta, s) = \alpha_{s}\cdot I_0\cdot\mathrm{e}^{-\int_{L(\theta, s)}\mu(\mathbf{x})\mathrm{d}\mathbf{x}},\quad \forall L(\theta, s)\in\mathbf{\Pi},
\end{equation}
where $\mathbf{\Pi}$ denotes the set of all X-rays, $L(\theta, s)$ is an X-ray received by an X-ray detector $s$ at a projection view $\theta$, and $I(\theta, s)$ is the number of received photons. $I_0$ is the number of photons emitted by an X-ray source. $\alpha_{s} \geq 0$ is the response factor of the detector $s$, measuring its detection ability. $\mu(\mathbf{x})$ represents the attenuation coefficient of observed objects at the position $\mathbf{x}$.
\par By assuming ideal X-ray detectors (\ie, $\alpha_{s}=1,\forall s$), the CT measurement data is given by:
\begin{equation}
        \rho(\theta, s)\triangleq-\ln\frac{I(\theta, s)}{\alpha\cdot I_0}= \int_{L(\theta, s)}\mu(\mathbf{x})\mathrm{d}\mathbf{x}.
        \label{eq:linear_model}
\end{equation}
Then, one can reconstruct high-quality CT images $\mu$ from the ideal measurement data $\rho$ using linear algorithms, such as FDK~\cite{fdk}. 
\par However, in practice, the responses of detectors are often non-ideal (\ie, $\alpha_s\neq 1,\forall s$) and difficult to maintain consistently (\ie, $\alpha_{s_1}\neq\alpha_{s_2}, \forall s_1, s_2\in S$, where $S$ denotes the set of the detectors) due to hardware limitations~\cite{boas2012ct}. The measurement can be defined by:
\begin{equation}
\begin{aligned}
        \widetilde{\rho}(\theta, s)&\triangleq-\ln\frac{I(\theta, s)}{I_0}\\ &=\begin{cases} 
-\ln{\alpha_{s}}+\int_{L(\theta, s)}\mu(\mathbf{x})\mathrm{d}\mathbf{x}, & \text{if}\ \alpha_{s} > 0\\ 
\ \texttt{NaN}, & \text{if}\ \alpha_{s} = 0.
\end{cases}
\end{aligned}
\label{eq:physical_effects}
\end{equation}
From Eq.~\ref{eq:physical_effects}, we observe two nonlinear physical effects: 
\begin{itemize}
    \item \textbf{Inconsistent Responses (IR)}, where non-ideal fluctuating (\ie, $\alpha_{s} > 0$ and $\alpha_{s} \neq 1$), inconsistent (\ie, $\alpha_{s_1}\neq\alpha_{s_2}, \forall s_1, s_2\in S$) detectors introduce an extra nonlinear term $-\ln\alpha_{s}$;
    \item \textbf{Invalid Measurements (IM)}, where the partial measurements by defective (\ie, $\alpha_{s} = 0$) detectors are invalid.
\end{itemize}
These two physical distortions make the inverse problem of reconstructing the unknown CT images $\mu$ from the real measurements $\widetilde{\rho}$ inherently nonlinear. Therefore, traditional linear algorithms, such as FDK~\cite{fdk}, are prone to severe ring artifacts in CT results.
\subsection{Problem Formulation}
\label{sec:problem_formulation}
\par Existing supervised DL methods~\cite{trapp2022deeprar, zamir2022restormer} mostly treat RAR as a post-processing denoising task. They first construct paired datasets $\{(\mu, \hat{\mu})_i\}_{i=1}^N$ by simulating artifact-corrupted images $\hat{\mu}$ from clean CT scans $\mu$, and train networks to learn a mapping $\mu = \phi(\hat{\mu})$. While effective in-domain, supervised learning lacks physical modeling of artifact formation, leading to poor generalization on OOD data. Moreover, their high memory demands limit applicability to 3D CBCT.
\par In contrast, we formulate RAR as a multi-parameter inverse problem grounded in CT physics. As highlighted in our theoretical analysis (Sec.~\ref{sec:theoritical_analyze}), the IR and IM effects are fundamental causes of CT ring artifacts. To address them, we introduce two vectors: $\boldsymbol{\alpha} = [\alpha_1,\dots,\alpha_s,\dots] \in \mathbb{R}^{|S|}$ to model detector response factors, and $\boldsymbol{\beta} = [\beta_1,\dots,\beta_s,\dots] \in \mathbb{R}^{|S|}$ to eliminate invalid measurements. Joint estimation of $\boldsymbol{\alpha}$ and $\boldsymbol{\beta}$ enables effective correction of both IR and IM, resulting in artifact-free reconstructions. Thus, our goal is to recover the image $\mu$ and estimate $(\boldsymbol{\alpha}, \boldsymbol{\beta})$ from measurement data $\widetilde{\rho}$ in an \textit{unsupervised} way. Formally, we aim to solve the following optimization problem:
\begin{equation}
    \begin{aligned}
        \mu^*,\ \boldsymbol{\alpha}^*,\ \boldsymbol{\beta}^*=\ & \underset{\mu,\ \boldsymbol{\alpha},\ \boldsymbol{\beta}}{\arg\min}\quad\mathcal{L}(\widetilde{\rho},\ \widehat{\rho}) \\
        &\mathrm{subject\ to}\quad\widehat{\rho} = \boldsymbol{A}(\mu,\ \boldsymbol{\alpha},\ \boldsymbol{\beta}) \\
        &\quad\quad\quad\quad\quad \boldsymbol{\alpha} \succeq \mathbf{0},\quad\boldsymbol{\beta} \in \{0,1\}^{|S|},
    \end{aligned}
    \label{eq:formulation}
\end{equation}
where $\mu^*,\ \boldsymbol{\alpha}^*,\ \boldsymbol{\beta}^*$ denote the underlying optimal solutions, $\widehat{\rho}$ represents estimated measurement, $\mathcal{L}$ is the loss function defined in Eq.~\eqref{eq:loss}, and $\boldsymbol{A}$ is a differentiable physical forward model that characterizes the CT acquisition process under the RAR setting, as detailed in Sec.~\ref{sec:phyiscal_forward_model}.
\paragraph{Parametrizing CT Images via INR} 
\par Due to the introduction of the physical parameters $(\boldsymbol{\alpha},\boldsymbol{\beta})$, this inverse problem (Eq.~\ref{eq:formulation}) is highly ill-posed, where multiple feasible solutions exist. Our Riner introduces INR to reconstruct the underlying CT image $\mu$. Specifically, we represent the CT image as a continuous function of spatial coordinates parametrized by an MLP network as follows:
\begin{equation}
    f_{\mathbf{\Phi}}:\mathbf{x}=(x,y,z)\in\mathbb{R}^3\rightarrow\mu(\mathbf{x})\in\mathbb{R},
    \label{eq:inr}
\end{equation}
where ${\mathbf{\Phi}}$ denote the learnable weights of the MLP network. Numerous studies~\cite {mildenhall2021nerf, tancik2020fourier, rahaman2019spectral, sitzmann2020implicit} have demonstrated that neural networks, when used to approximate images, typically capture low-frequency global structures first and then gradually recover high-frequency details. We aim to leverage this inherent learning prior to constrain the solution space of the ill-posed inverse problem (Eq.~\ref{eq:formulation}), resulting in excellent CT reconstructions. 
\subsection{Differential Physical Forward Model}
\label{sec:phyiscal_forward_model}
\par Following the problem formulation in Eq.~\eqref{eq:formulation} and Eq.~\eqref{eq:inr}, we optimize the MLP network $f_\mathbf{\Phi}$ to map spatial coordinates $\mathbf{x}$ to CT intensities $\mu(\mathbf{x})$. To achieve this, we introduce a new \textit{differential} physical forward model, which converts the MLP-predicted intensities $\mu(\mathbf{x})=f_\mathbf{\Phi}(\mathbf{x})$ and the variables $(\alpha_s,\beta_s)$ into estimated measurements $\widehat{\rho}(\theta,s)$, while allowing for gradient back-propagation. Building on the theoretical results in Eq.~\eqref{eq:physical_effects}, our forward model $\boldsymbol{A}$ is defined as follows:
\begin{equation}
    \begin{aligned}
            \boldsymbol{A}:\quad\widehat{\rho}&(\theta,s) =  \Bigl[-\ln\alpha_s + \sum_{\mathbf{x}\in L(\theta, s)}\mu(\mathbf{x}) \cdot \Delta \mathbf{x}\Bigr] \cdot \beta_s, \\
        &\mathsf{with}\quad \alpha_s = \max(\alpha_s^0,\ \epsilon), \quad\beta_s = \sigma(\beta_s^0),
        % \widehat{\rho}&(\theta,s) = \boldsymbol{A}(\mu, \alpha_s, \beta_s) \triangleq \Bigl[-\ln\alpha_s + \sum_{\mathbf{x}\in L(\theta, s)}\mu(\mathbf{x}) \cdot \Delta \mathbf{x}\Bigr] \cdot \beta_s, \\
        % &\mathrm{with}\quad\mu(\mathbf{x})=f_\mathbf{\Phi}(\mathbf{x}),\quad\alpha_s = \mathsf{Max}(\alpha_0, \epsilon), \quad\beta_s = \mathsf{Sigmoid}(\beta_0),
    \end{aligned}
    \label{eq:physical_model}
\end{equation}
where $\alpha_s^0$ is initialized to 1, and $\epsilon = 1\times10^{-8}$ is a small constant. These ensure the non-negativity of the response factor $\alpha_s$ and the validity of the logarithmic computation. $\Delta\mathbf{x} = \|\mathbf{x}_a - \mathbf{x}_b\|_2$ represents the Euclidean distance between adjacent coordinates $\mathbf{x}_a, \mathbf{x}_b \in L(\theta, s)$ along the X-ray path. While the variable $\beta_s$ acts as a binary mask, where $\beta_s = 0$ for defective detectors and $1$ otherwise. It is generated using the $\mathsf{Sigmoid}$ function with an initial value of $\beta_s^0 = 1$.
\par Superior to the conventional linear integral model (Eq.~\ref{eq:linear_model}) ~\cite{shen2022nerp,zha2022naf,wu2023self,zang2021intratomo,lin2023learning}, our model effectively addresses the CT RAR problem by introducing additional physical parameters for more accurate acquisition modeling. Specifically, it includes \textbf{1) Paramerization of the detector responses}, where our model explicitly estimates the response factor $\alpha_s$ to alleviate the IR effect; \textbf{2) Removal of invalid measurements}, where we learn the mask $\beta_s$ to block the gradient back-propagation from invalid signals measured by defective X-ray detectors, thereby eliminating the IM effect.
\subsection{Model Optimization}
\label{sec:model_optimization}
\par Fig.~\ref{fig:fig_method} shows the ray-based optimization pipeline for our Riner model. We first sample a set of spatial coordinates $\mathbf{x}$ along any X-ray $L(\theta, s)$ at a pre-defined and fixed interval $\Delta\mathbf{x}$. Then, we feed these coordinates into the MLP network $f_\mathbf{\Phi}$ to predict the corresponding CT intensities $\mu(\mathbf{x})$. Furthermore, we conduct our differentiable forward model $\boldsymbol{A}$ (Eq.~\ref{eq:physical_model}) to convert these MLP-predicted intensities $\mu(\mathbf{x})=f_\mathbf{\Phi}(\mathbf{x}),\forall \mathbf{x}\in L(\theta, s)$, the corresponding learnable response factor $\alpha_s$, and the mask $\beta_s$ into estimated measurement $\widehat{\rho}(\theta, s)$. Finally, the MLP network $f_\mathbf{\Phi}$ and the parameters $(\alpha_s, \beta_s)$ are jointly optimized using gradient descent backpropagation algorithms, such as Adam~\cite{kingma2014adam}, to minimize the loss function $\mathcal{L}$, which is defined by:
\begin{equation}
\begin{aligned}
    \mathcal{L}(\widetilde{\rho},\ \widehat{\rho}) \triangleq 
\underbrace{
\sum_{L(\theta, s)\in\bar{\mathbf{\Pi}}}\big\|\widehat{\rho}(\theta,s)-\widetilde{\rho}(\theta,s)\cdot\beta_s\big\|_1
}_{\text{Data Consistency}} + \underbrace{\lambda\cdot\sum_{s\in \bar{S}}-\beta_s^2}_{\text{Negative } \ell_2}\ ,
\end{aligned}
\label{eq:loss}
\end{equation}
where $\bar{\mathbf{\Pi}} \subseteq \mathbf{\Pi}$ is a random subset of X-rays, and $\bar{S} \subseteq S$ is the corresponding detector subset. The data consistency minimizes the discrepancy between the predicted and actual measurements, while the negative $\ell_2$ regularizes the masks $\beta_s$, preventing all of them from converging to zero (\ie, identifying all detectors as defective), thereby avoiding optimization degradation. $\lambda = 0.01$ is a hyperparameter that controls the contribution of the regularization.
\par Fortunately, \textit{this ray-based optimization is memory-friendly}, as each iteration only involves forward and backward passes over a small set of sampled voxels, rather than the entire 3D volume as in supervised methods~\cite{hybridcnn, trapp2022deeprar, zamir2022restormer}. This enables our approach to scale seamlessly to CBCT scans. After model optimization, the MLP network $f_{{\mathbf{\Phi}}^*}$ and the vectors $\boldsymbol{\alpha}^*,\boldsymbol{\beta}^*$ represent the desired CT image, detector responses, and defective detectors, respectively. We finally reconstruct the high-quality CT image $\mu^*$ by feeding all coordinates $\mathbf{x}$ into the well-optimized network $f_{{\mathbf{\Phi}}^*}$.
\begin{table*}[t]
\centering
\setlength{\tabcolsep}{1.5mm}
% \resizebox{\textwidth}{!}{
\begin{tabular}{clcccccc} 
\toprule

\multirow{2.5}{*}{\textbf{Category}}     & \multirow{2.5}{*}{\textbf{Method}} & \multicolumn{2}{c}{\textbf{DeepLesion}}   & \multicolumn{2}{c}{\textbf{LIDC}}   & \multicolumn{2}{c}{\textbf{AAPM}}       \\ 
\cmidrule(r){3-4}\cmidrule(r){5-6}\cmidrule(r){7-8}
                              &                         & PSNR           & SSIM            & PSNR           & SSIM & PSNR           & SSIM            \\ 
\midrule
\multirow{4}{*}{\texttt{Model-based}}  & FBP/FDK                     & 13.81$\pm$3.01 & 0.505$\pm$0.129                & 12.29$\pm$2.86 & 0.452$\pm$0.153 & 15.26$\pm$1.24 & 0.695$\pm$0.062 \\
& Norm                     & 24.37$\pm$1.95 & 0.774$\pm$0.067                & 24.61$\pm$2.31 & 0.786$\pm$0.049  & 26.86$\pm$0.50 & 0.865$\pm$0.024\\
                              & WaveFFT                   & 20.70$\pm$2.48 & 0.663$\pm$0.096                & 19.44$\pm$2.22 & 0.622$\pm$0.118                & 23.83$\pm$1.17 & 0.777$\pm$0.051 \\ 
                              & Super~                   & 29.11$\pm$3.43 & 0.815$\pm$0.084                & 31.24$\pm$3.14 & 0.855$\pm$0.051               & 30.06$\pm$3.64 & 0.915$\pm$0.026  \\ 

\midrule
\multirow{5}{*}{\texttt{Supervised}}             & HyUNet                & 33.73$\pm$1.88 & 0.893$\pm$0.035                & 32.46$\pm$1.54 & 0.854$\pm$0.051     & 31.49$\pm$1.58 & 0.871$\pm$0.041\\
& DeepRAR                & 34.61$\pm$2.13 & 0.893$\pm$0.043                & 30.34$\pm$3.54 & 0.802$\pm$0.085                 & 28.35$\pm$1.61 & 0.821$\pm$0.053\\
& NAFNet                 & 36.51$\pm$1.91 & 0.937$\pm$0.030               & 36.07$\pm$1.91 & 0.917$\pm$0.051               & 33.89$\pm$1.69 & 0.907$\pm$0.034 \\ 
                              & Restormer                 & \underline{37.31$\pm$1.92} & \underline{0.947$\pm$0.028}               & \underline{36.69$\pm$1.90} & \underline{0.925$\pm$0.051}               & \underline{33.94$\pm$1.50} & \underline{0.907$\pm$0.031} \\ 
& MambaIRv2                 & 36.77$\pm$1.88 & 0.939$\pm$0.029               & 35.83$\pm$1.87 & 0.912$\pm$0.050               & 33.23$\pm$2.03 & 0.901$\pm$0.037 \\ 
\midrule
\multirow{2}{*}{\texttt{Unsupervised}} 

% &NAF            & 35.09$\pm$2.17 & 0.907$\pm$0.033                & 35.01$\pm$1.57 & 0.898$\pm$0.045& 34.12$\pm$0.97 & 0.916$\pm$0.026\\
&SinoRAR            & 32.70$\pm$1.77 & 0.881$\pm$0.034                & 32.63$\pm$1.21 & 0.879$\pm$0.035& \texttt{N/A}               & \texttt{N/A}\\
& Riner (Ours)                    & \textbf{39.02$\pm$2.18} & \textbf{0.967$\pm$0.019}                 & \textbf{39.11$\pm$2.17} & \textbf{0.962$\pm$0.030} & \textbf{36.53$\pm$1.28} & \textbf{0.949$\pm$0.020}\\
\bottomrule
\end{tabular}
\caption{Quantitative results of our Riner and compared methods on three simulated datasets, including 2D FBCT DeepLesion~\cite{deeplesion}, 2D FBCT LIDC~\cite{lidc}, and 3D CBCT AAPM~\cite{aapm}. The best and second performances are highlighted in \textbf{blod} and \underline{underline}, respectively.}
\label{table_com}
% }
\end{table*}
\section{Experiments}
% \par In this section, we conduct extensive experiments to assess our unsupervised method, including 1) comparisons (Sec.~\ref{sec:Comparison}) with SOTA RAR approaches on both simulated and real-world data, and 2) ablation studies on the key components of our model (Sec.~\ref{sec:Ablation Studies}).
\subsection{Experimental Settings}
\label{sec:exp_setting}
\paragraph{Datasets}
\par We perform experiments on three simulated datasets and two real-world datasets as follows: \textbf{1) Simulated 2D FBCT DeepLesion}~\cite{deeplesion}, consisting of 5,261 2D slices of a 256$\times$256 size from multiple 3D volumes. We split these 2D slices into three parts: 5,061 slices for the training set, 100 slices for the validation set, and 100 slices for the test set. \textit{Notably, our unsupervised Riner does not access the training and validation sets used for training supervised baselines}; \textbf{2) Simulated 2D FBCT LIDC}~\cite{lidc}, consisting of 100 2D slices of a 256$\times$256 size from multiple 3D volumes for an additional test set; \textbf{3) Simulated 3D CBCT AAPM}~\cite{aapm}, consisting of 10 3D volumes, each with sizes of 256$\times$256$\times$100, used to evaluate our Riner model and baselines within 3D CBCT geometry; \textbf{4) Real-world 2D FBCT Walnut}, consisting of a 2D slice with sizes of 512$\times$512 and a physical resolution of 60$\times$60 $\mu$m$^\text{2}$, acquired by a commercial Bruker SKYSCAN 1276 micro-CT scanner using 2D FBCT geometry; \textbf{5) Real-world 3D CBCT Chicken foot}, consisting of a 3D volume with sizes of 512$\times$512$\times$80 and a physical resolution of 60$\times$60$\times$60 $\mu$m$^\text{3}$, acquired by a commercial Bruker SKYSCAN 1276 micro-CT scanner using 3D CBCT geometry. \textit{We provide details on data pre-processing in the Appendix.}
\paragraph{Baselines and Metrics}
\par We compare Riner against 10 cutting-edge methods, including 1) 4 model-based algorithms (FBP/FDK~\cite{fbp,fdk}, Norm~\cite{rivers1998Norm}, WaveFFT~\cite{wavefft}, and Super~\cite{vo2018Super}); 2) 5 supervised DL models (HyUNet~\cite{hybridcnn}, DeepRAR~\cite{trapp2022deeprar}, NAFNet~\cite{chen2022simple}, Restormer~\cite{zamir2022restormer}, and MambaIRv2~\cite{guo2024mambair,guo2024mambairv2}); 3) 1 unsupervised DL model (SinoRAR~\cite{SinoRAR}). \textit{The five supervised methods are trained using the training and validation sets of the DeepLesion dataset~\cite{deeplesion}}. Moreover, we use PSNR and SSIM as quantitative metrics. \textit{More implementation details about the baselines can be found in the Appendix.}
\paragraph{Implementation Details}
\par We implement our Riner using PyTorch~\cite{paszke2019pytorch}. The network $f_{\mathbf{\Phi}}$ consists of a hash encoding module~\cite{muller2022instant} followed by two fully connected (FC) layers, with a ReLU activation applied after the first FC layer. The hash encoding accelerates optimization and enhances the reconstruction of high-frequency details. We adopt the hyperparameters suggested in the original paper~\cite{muller2022instant}: $L=10$, $T=2^{10}$, $F=8$, $N_\text{min}=2$, and $b=2$. During each step, we randomly sample a subset of 80 X-rays (\ie, $|\bar{\mathbf{\Pi}}| = 80$) from 2 detectors (\ie, $|\bar{S}| = 2$) across 40 projection views. We use the Adam optimizer~\cite{kingma2014adam} with default settings. The learning rate is set to $10^{-3}$ and decays by a factor of 0.5 every 1,000 iterations over a total of 4,000 iterations. The optimization is performed on a single NVIDIA RTX 4090 GPU, requiring only around 3 GB of memory. \textit{All hyperparameters of Riner are tuned using 10 samples from the DeepLesion dataset~\cite{deeplesion} and remain unchanged in all other samples.}
\begin{figure}[t]
    \centering
    \includegraphics[width=0.3\textwidth]{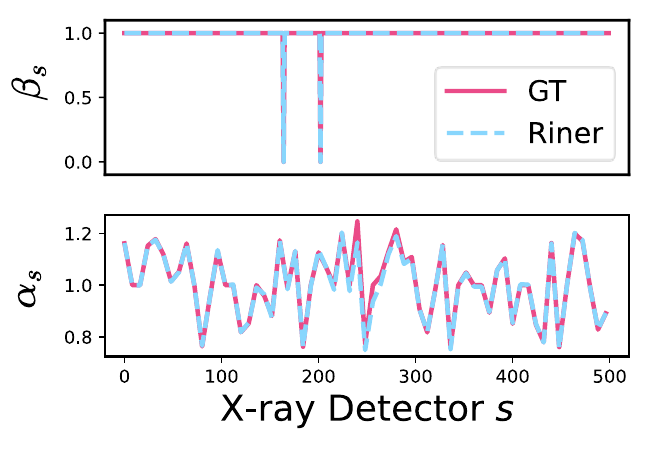}  
    \caption{Parameter estimations of our Riner on a sample ($\#$9) of DeepLesion dataset~\cite{deeplesion}.}
    \label{fig:fig_etimated_responses}
\end{figure}
\begin{figure*}[t]
    \centering
    \includegraphics[width=0.8\textwidth]{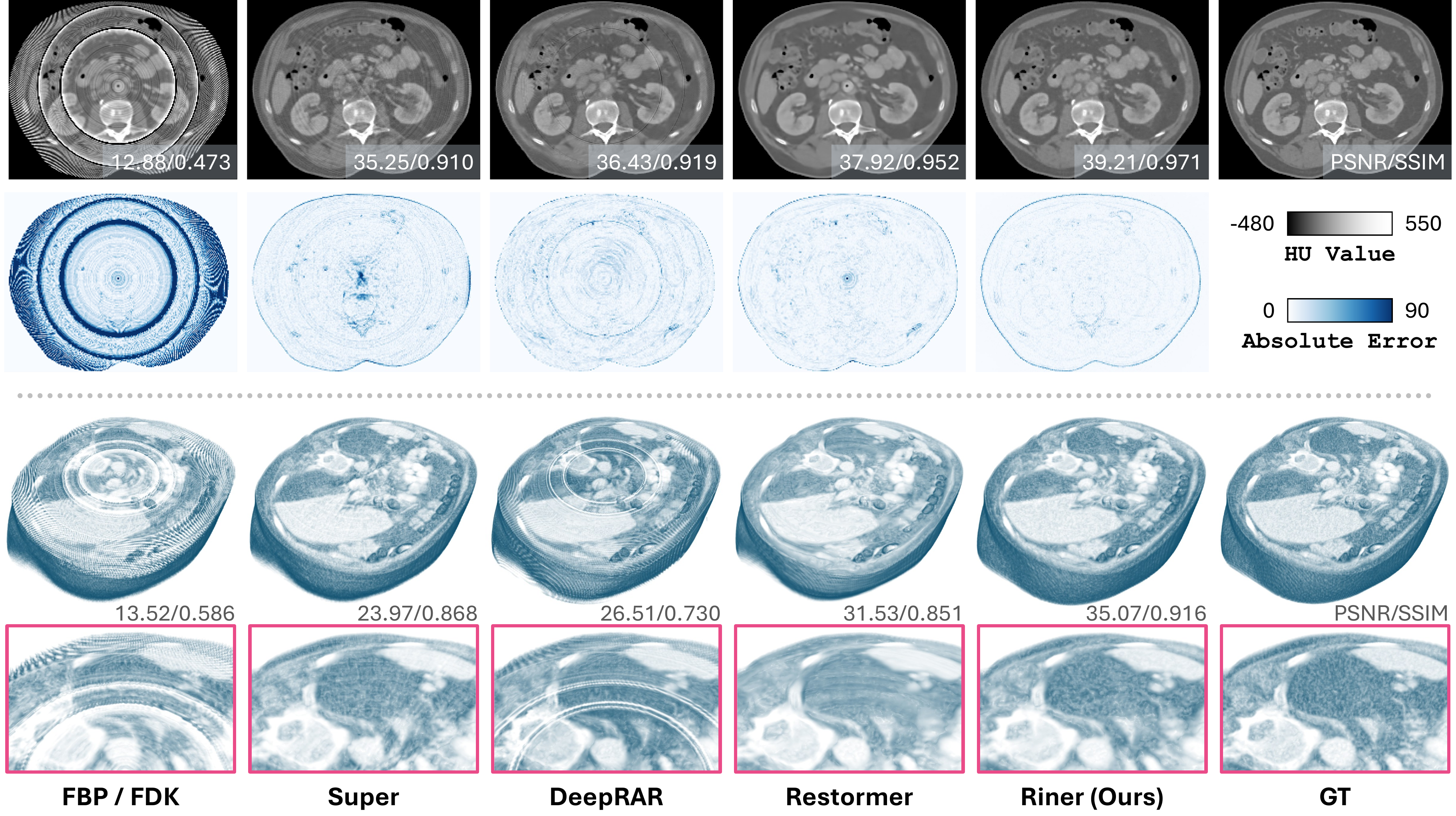}
    \caption{Qualitative results of our Riner and compared methods on two representative samples of three simulated datasets (2D FBCT DeepLesion~\cite{deeplesion} and 3D CBCT AAPM~\cite{aapm}).}
    \label{fig:fig_com_2d}
\end{figure*}
\subsection{Main Results}
\label{sec:Comparison}
\paragraph{Comparisons on Simulated Datasets}
% \par We first compare our Riner with 10 baselines on three simulated datasets. Note that the five supervised methods (HyUnet, DeepRAR, NAFNet, Restormer, and MambaIRv2) are specifically designed for 2D images. We apply these 2D models slice-by-slice to reconstruct 3D volumes. Since SinoRAR operates on 2D measurement data, it is excluded from the comparisons on the 3D AAPM dataset. 
\par Table~\ref{table_com} shows the quantitative results. Riner consistently achieves the best performance across all three simulated datasets, surpassing the best-performing supervised baseline, Restormer, by $+$1.71 dB, $+$2.42 dB, and $+$2.59 dB in PSNR, respectively. Among model-based methods, Super achieves the best numerical results due to its modeling of multiple types of ring artifacts. Among the supervised baselines, NAFNet, Restormer, and MambaIRv2 perform well on 2D datasets due to their advanced architectures, but degrade notably on the 3D data. In contrast, our Riner maintains superior and stable RAR performance across both 2D and 3D datasets, owing to its physics-informed formulation.  
\par The core idea of our Riner for addressing the CT RAR problem is to explicitly estimate the detector physical parameters $(\boldsymbol{\beta},\boldsymbol{\alpha})$. Thus, accurately estimating these parameters $(\boldsymbol{\beta},\boldsymbol{\alpha})$ is critical for enabling reliable CT reconstructions. In Fig.~\ref{fig:fig_etimated_responses}, we visualize the parameters estimated by our Riner alongside GTs on a sample of the DeepLesion dataset. These results confirm that Riner can accurately estimate the physical parameters, enabling reliable CT results.
\par Fig.~\ref{fig:fig_com_2d} demonstrates qualitative comparisons. Visually, our Riner consistently reconstructs clear, high-fidelity CT images with fine structures, closely matching the GTs and significantly outperforming all baselines. \textit{More visual comparisons are provided in the Appendix.}
\paragraph{Comparisons on Real-World Datasets}
\par
\begin{figure}[t]
    \centering
    \includegraphics[width=\linewidth]{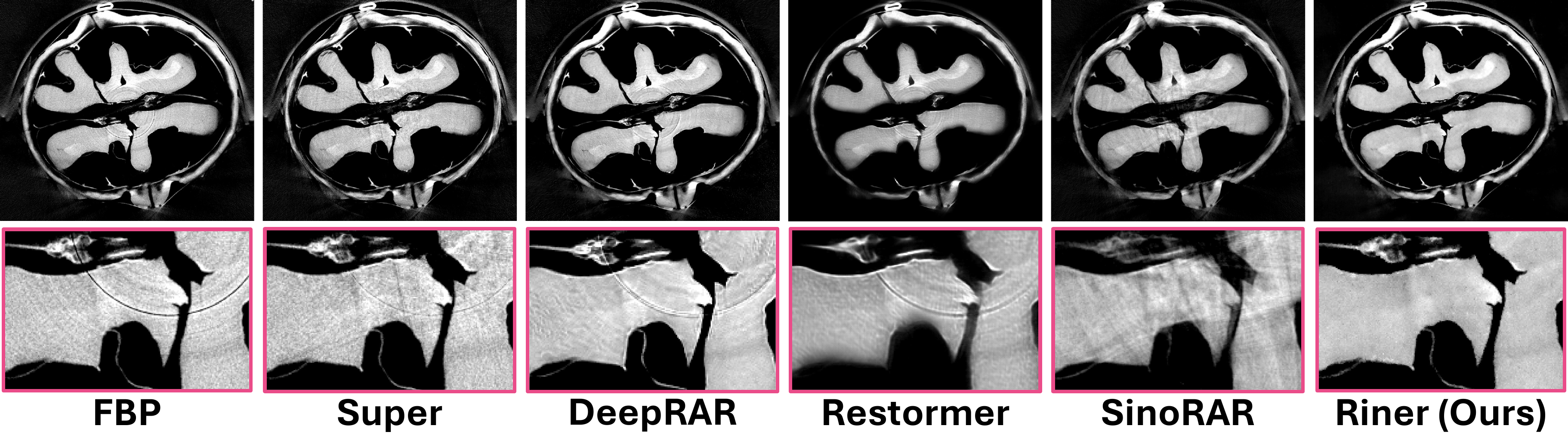}
    \caption{Qualitative results of our Riner and compared methods on a real-world \textit{Walnut} sample (see Appendix) with sizes of 512$\times$512 and an ultra-high resolution of 60$\times$60 $\mu\text{m}^2$, acquired by a commercial Bruker SKYSCAN 1276 micro-CT scanner.}
    \label{fig:fig_com_2d_real}
\end{figure}

 Micro-CT imaging is widely used in fields such as biomedical research and materials science~\cite{ritman2011current}. However, ring artifacts are common in Micro-CT due to its $\mu$m-scale ultra-high physical resolution~\cite{yousuf2010efficient, rashid2012improved}. Here, we validate the effectiveness of our Riner on real-world Micro-CT imaging. We scan a \textit{Walnut} sample and a \textit{Chicken foot} sample using a commercial Bruker SKYSCAN 1276 micro-CT scanner under 2D FBCT and 3D CBCT protocols (see Appendix). We only report qualitative results since artifact-free GT images are difficult to obtain. As demonstrated in Fig.~\ref{fig:3d_real} and Fig.~\ref{fig:fig_com_2d_real}, our Riner effectively removes ring artifacts while preserving image details, significantly outperforming baselines. 
% The hyperparameters for our method remain consistent across simulated and real-world data, demonstrating its robustness.
\begin{figure}[t]
    \centering
    \includegraphics[width=0.8\linewidth]{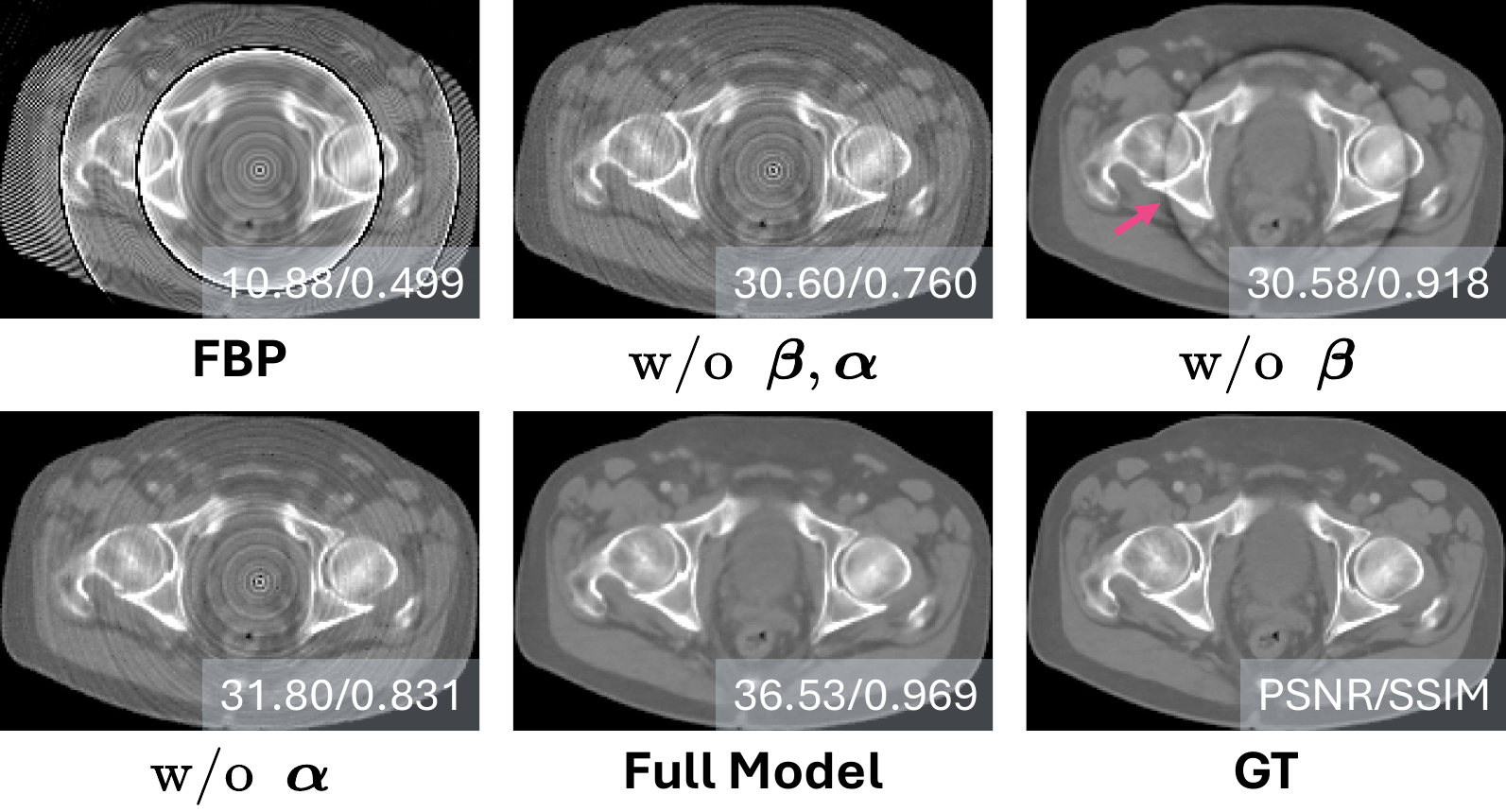}
    \caption{Qualitative results of our Riner ablating the physical forward model (Eq.~\ref{eq:physical_model}) on a representative sample ($\#$6) of DeepLesion dataset~\cite{deeplesion}.}
    \label{fig:fig_com_ab_forward}
\end{figure}
\begin{table}[t]
    \centering
    \begin{tabular}{cccc}
    \toprule
     \multicolumn{2}{c}{\textbf{Physical Model}}    & \multirow{2.5}{*}{PSNR}& \multirow{2.5}{*}{SSIM}\\ \cmidrule(r){1-2}
     Mask $\boldsymbol{\beta}$& Factor $\boldsymbol{\alpha}$ & &\\ \midrule
     \ding{55} & \ding{55} &35.34$\pm$2.26 &0.890$\pm$0.050\\
     \ding{55} & \ding{51} & 31.05$\pm$9.16 & 0.894$\pm$0.130\\ 
     \ding{51} & \ding{55} & 35.72$\pm$1.93& 0.910$\pm$0.033\\
     \ding{51} & \ding{51} & \textbf{38.98$\pm$1.41}&\textbf{0.974$\pm$0.005}\\ \bottomrule
    \end{tabular}
    \caption{Quantitative results of our Riner ablating the physical forward model (Eq.~\ref{eq:physical_model}) on 10 samples of DeepLesion dataset~\cite{deeplesion}. }
    \label{tab:table_ab_forward}
\end{table}
% \begin{figure}[t]
%   \centering
%   \begin{minipage}{0.55\linewidth}
%     \centering
%     \includegraphics[width=\linewidth]{fig/fig_com_ab_forward.pdf}
%     \caption{Qualitative results of our Riner ablating the physical forward model (Eq.~\ref{eq:physical_model}) on a sample ($\#$6) of DeepLesion dataset~\cite{deeplesion}.}
%     \label{fig:fig_com_ab_forward}
%   \end{minipage}%
%   \hspace{0.04\linewidth}
%     \begin{minipage}{0.35\linewidth}
%         \captionof{table}{Quantitative results of our Riner ablating the physical forward model (Eq.~\ref{eq:physical_model}) on 10 samples of DeepLesion dataset~\cite{deeplesion}. }
%         \resizebox{\linewidth}{!}{
%         \begin{tabular}{ccc}
%         \toprule
%          \multicolumn{2}{c}{\textbf{Physical Model}}    & \multirow{2.5}{*}{PSNR}\\ \cmidrule(r){1-2}
%          Mask $\boldsymbol{\beta}$& Factor $\boldsymbol{\alpha}$ & \\ \midrule
%          \ding{55} & \ding{55} &35.34$\pm$2.26 \\
%          \ding{51} & \ding{55} & 31.05$\pm$9.16 \\ 
%          \ding{55} & \ding{51} & 35.72$\pm$1.93\\
%          \ding{51} & \ding{51} & \textbf{38.98$\pm$1.41}\\ \bottomrule
%         \end{tabular}}
%         \label{table_ab_forward}
%   \end{minipage}
% \end{figure}
\begin{figure}[t]
    \centering
    \includegraphics[width=\linewidth]{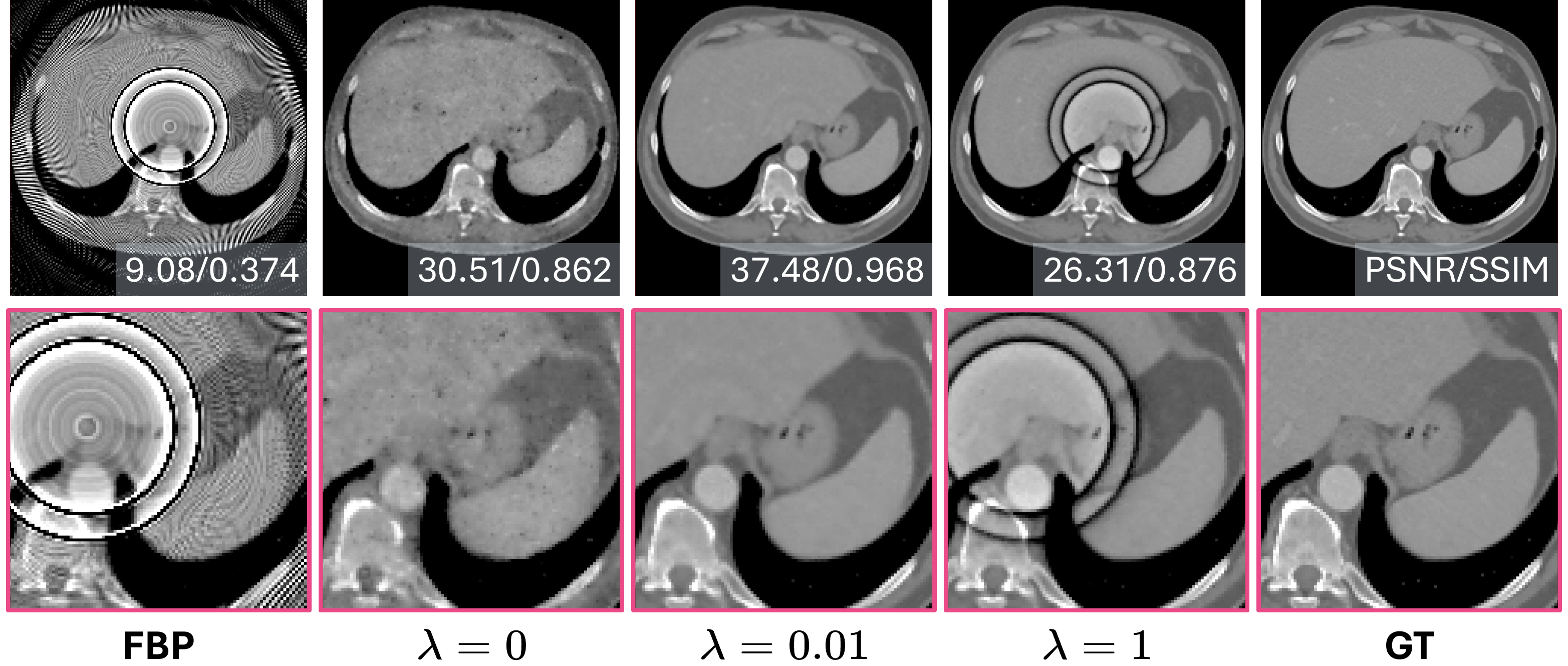}
    \caption{Qualitative results of our Riner ablating the negative $\ell_2$ term (Eq.~\ref{eq:loss}) on a representative sample ($\#$3) of DeepLesion dataset~\cite{deeplesion}.}
    \label{fig:fig_ablation_lambda}
\end{figure}
\begin{table}[t]
    \centering
    \begin{tabular}{lcc} 
    \toprule
    Weight & PSNR & SSIM \\ \midrule
    $\lambda = 0$ & 31.63$\pm$1.77 &  0.881$\pm$0.034\\ 
    $\lambda = 0.01$ & \textbf{38.98$\pm$1.41} & \textbf{0.974$\pm$0.005}\\
    $\lambda = 1$ & 31.55$\pm$8.97 & 0.896$\pm$0.134\\
    \bottomrule
    \end{tabular}
    \caption{Quantitative results of our Riner ablating the negative $\ell_2$ term (Eq.~\ref{eq:loss}) on 10 samples of DeepLesion dataset~\cite{deeplesion}.}
    \label{table:ablation_n}
\end{table}
% \begin{figure}[t]
%      \centering
%   \begin{minipage}{0.725\linewidth}
%     \centering
%     \includegraphics[width=\linewidth]{fig/fig_ablation_lambda.pdf}
%     \caption{Qualitative results of our Riner ablating the negative $\ell_2$ term (Eq.~\ref{eq:loss}) on a sample ($\#$3) of DeepLesion dataset~\cite{deeplesion}.}
%     \label{fig:fig_ablation_lambda}
%   \end{minipage}%
%     \hfill
%     \begin{minipage}{0.25\linewidth}
%         \centering
%         \captionof{table}{Quantitative results of our Riner ablating the negative $\ell_2$ term (Eq.~\ref{eq:loss}) on 10 samples of DeepLesion dataset~\cite{deeplesion}.}
%         \label{table:ablation_n}
%         \resizebox{\linewidth}{!}{
%         \begin{tabular}{lc} 
%         \toprule
%         Weight & PSNR \\ \midrule
%         $\lambda = 0$ & 31.63$\pm$1.77\\ 
%         $\lambda = 0.01$ & \textbf{38.98$\pm$1.41}\\
%         $\lambda = 1$ & 31.55$\pm$8.97\\
%         \bottomrule
%         \end{tabular}}
%   \end{minipage}
% \end{figure}

\subsection{Ablation Studies}
\label{sec:Ablation Studies}
\paragraph{Influence of Physical Forward Model}
\par The proposed forward model $\boldsymbol{A}$ introduce the learnable variables $(\boldsymbol{\beta},\boldsymbol{\alpha})$ to correct the IM and IR effects, respectively. Here, we investigate their effectiveness. Fig.~\ref{fig:fig_com_ab_forward} and Table~\ref{tab:table_ab_forward} show the qualitative and quantitative results. There are four cases: 
\begin{itemize}
    \item \underline{w/o $\boldsymbol{\beta}, \boldsymbol{\alpha}$ (Integral Model):} \ie, degrading our physical model to the traditional integral model in Eq.~\eqref{eq:linear_model}. We observe that although the integral model reduces some ring artifacts using the continuous prior of INR, it still contains significant artifacts due to the IM and IR effects.
    \item \underline{w/o $\boldsymbol{\beta}$:} When removing the mask $\boldsymbol{\beta}$, our Riner fails to handle the IM effect. The image contains black ring artifacts of the IM effect, indicated by \textbf{the pink arrow} in Fig.~\ref{fig:fig_com_ab_forward}. While the artifacts of the IR effect are effectively reduced by estimating the response $\boldsymbol{\alpha}$.
    \item \underline{w/o $\boldsymbol{\alpha}$:} This case is opposite to case 2. The IM effect artifacts are removed, but the IR effect artifacts remain.
    \item \underline{Full Model:} Our Riner method can effectively eliminate both types of ring artifacts.
\end{itemize}
Generally, the results of ablating the physical model align perfectly with the expectations of our theoretical modeling for the IR and IM effects, showing its effectiveness.
\paragraph{Influence of Negative $\ell_2$ Regularization}
\par Riner explicitly learns the mask $\boldsymbol{\beta}$ to remove signals from defective detectors, thereby addressing the IM effect. To avoid the trivial solution of $\boldsymbol{\beta} = \mathbf{0}$, we incorporate a negative $\ell_2$ regularization term into the loss $\mathcal{L}$ (Eq.~\ref{eq:loss}).  We evaluate its effect by testing different weights $\lambda \in \{0,0.01,1\}$. Fig.~\ref{fig:fig_ablation_lambda} and Table~\ref{table:ablation_n} show both qualitative and quantitative results. Without regularization (\ie, $\lambda=0$), the optimization quickly collapses to $\boldsymbol{\beta} = \mathbf{0}$, failing to reconstruct detailed CT images. In contrast, with a strong regularization of $\lambda=1$, the mask $\boldsymbol{\beta}$ converges to $\boldsymbol{1}$, meaning no defective detectors are identified. Therefore, the black ring artifacts caused by IM effects remain. The best performance is achieved at $\lambda = 0.01$, which achieves a balanced regularization. \textit{Based on this ablation study, we set $\lambda = 0.01$ as the default in all experiments.}
\paragraph{Influence of Network Architecture}
\par
\begin{figure}[!]
    \centering
    \includegraphics[width=0.8\linewidth]{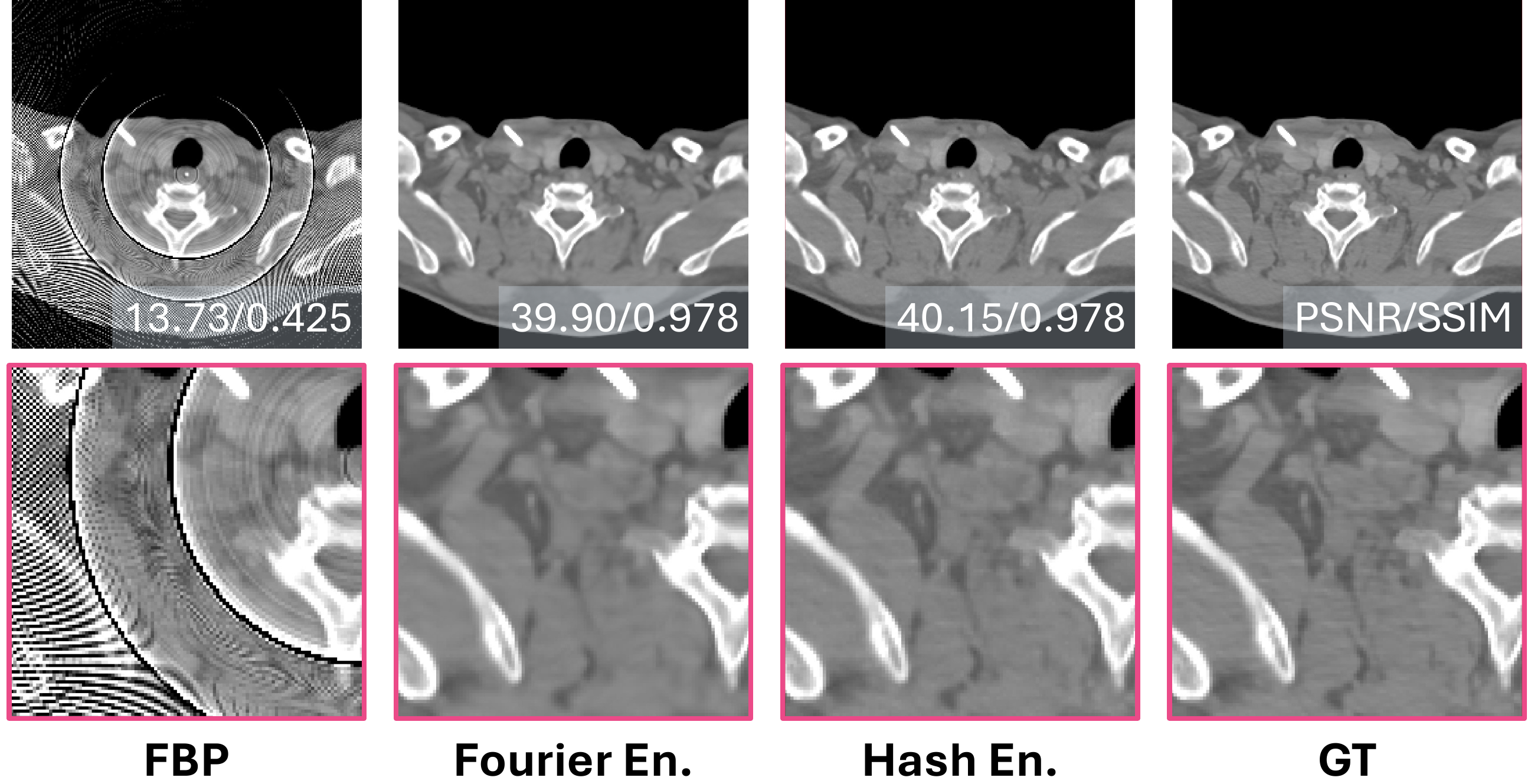}
    \caption{Qualitative results of our Riner ablating the network $f_\mathbf{\Phi}$ on a representative sample ($\#$4) of DeepLesion dataset~\cite{deeplesion}.}
    \label{fig:fig_ablation_net_img}
\end{figure}

 The network $f_\mathbf{\Phi}$ consists of hash encoding~\cite{muller2022instant} and a 2-layer MLP. Here, we study the influence of network architecture on RAR reconstruction. Specifically, we implement an alternative using Fourier encoding~\cite{tancik2020fourier} followed by a 4-layer MLP, while keeping all other settings identical for a fair comparison. Fig.~\ref{fig:fig_ablation_net_img} shows the reconstructed CT images. These results demonstrate that both networks achieve excellent performance, with the hash encoding yielding a slight improvement of $+$0.25 dB in PSNR (40.15 vs. 39.90). This suggests that the superior RAR performance of our Riner model is mainly attributed to the proposed physical modeling that effectively captures the nonlinear IM and IR effects, rather than the network architecture itself. 
% , and Fig.~\ref{fig:fig_ablation_net_para} presents the estimated parameters.
% \begin{figure}[t]
%     \centering
%     \includegraphics[width=0.3\textwidth]{fig/fig_ablation_net_para.pdf}  % Adjust the width as needed
%     \caption{Paramter estimations of our Riner ablating the network $f_\mathbf{\Phi}$ on a sample ($\#$4) of DeepLesion dataset~\cite{deeplesion}}
%     \label{fig:fig_ablation_net_para}
% \end{figure}

% \begin{figure}[t]
%   \centering
%   \begin{minipage}{0.6\linewidth}
%     \centering
%     \includegraphics[width=\linewidth]{fig/fig_ablation_net_img.pdf}
%     \caption{Image reconstructions of our Riner ablating the network $f_\mathbf{\Phi}$ on a sample ($\#$4) of DeepLesion dataset~\cite{deeplesion}.}
%     \label{fig:fig_ablation_net_img}
%   \end{minipage}%
%     \hspace{0.03\linewidth}
%     \begin{minipage}{0.325\linewidth}
%     \centering
%     \includegraphics[width=\linewidth]{fig/fig_ablation_net_para.pdf}  % Adjust the width as needed
%     \caption{Paramter estimations of our Riner ablating the network $f_\mathbf{\Phi}$ on a sample ($\#$4) of DeepLesion dataset~\cite{deeplesion}}
%     \label{fig:fig_ablation_net_para}
%   \end{minipage}
% \end{figure}
\section{Conclusion and Limitation}
\label{sec:conclusion}
\par In this paper, we present Riner, a novel approach to address the RAR problem in 3D X-ray CBCT. Unlike existing end-to-end RAR methods, our approach formulates RAR as a multi-parameter inverse problem, targeting the physical origins of ring artifact formation. Using a differentiable physical model, Riner jointly learns the neural representation of the artifact-free CT image and estimates the detector response parameters directly from raw measurements. 
% Extensive experiments on five datasets demonstrate that our unsupervised method achieves SOTA RAR performance.'
\par While effective, the efficiency of our method can be further improved. As an unsupervised approach, Riner requires case-specific optimization, taking about 30 seconds for a 2D slice of 256$\times$256 and 10 minutes for a 3D volume of 256$\times$256$\times$100 on an NVIDIA RTX 4090 GPU. Advanced techniques (\eg, 3D GS) may improve its efficiency.
% \par Despite effectiveness, the efficiency of our method could be further improved. As an unsupervised technique, Riner requires case-specific optimization. It takes approximately 30 seconds for a 2D slice of 256$\times$256 and 10 minutes for a 3D volume of 256$\times$256$\times$100 on an NVIDIA RTX 4090 GPU. Advanced acceleration techniques (\eg, meta learning) may offer promising solutions to improve its efficiency.
\section{Acknowledgment}
\par This work was supported by the National Natural Science Foundation of China under Grant No. 62571328.
\bibliography{aaai2026}
\section{Appendix}
\subsection{Details of Data Pre-processing}
\par In this work, we conduct extensive experiments on five datasets to evaluate our method. For the three simulated datasets (DeepLesion~\cite{deeplesion}, LIDC~\cite{lidc}, and AAPM~\cite{aapm}), we follow the data simulation protocols established in prior RAR studies~\cite{trapp2022deeprar,hybridcnn,vo2018Super} to generate CT ring artifacts. We consider three types of detectors: ideal detectors with full signal response (\ie, $\alpha_s=1$), non-ideal detectors with fluctuating responses (\ie, $\alpha_s>0$ and $\alpha_s\neq1$), and defective detectors with no signal response (\ie, $\alpha_s=0$). Specifically, 75\% of detectors are randomly set as non-ideal with response factors $\alpha_s$ sampled from [0.75, 1.25], and two detectors are randomly set as defective. The number of emitted photons is set to 1$\times$10\textsuperscript{7}. We implement equidistant 2D FBCT and 3D CBCT protocols using the Python library \texttt{CIL}~\cite{cil1,cil2}, and incorporate Poisson noise into the measurements.
\par
\begin{figure}[t]
    \centering
    \includegraphics[width=0.98\linewidth]{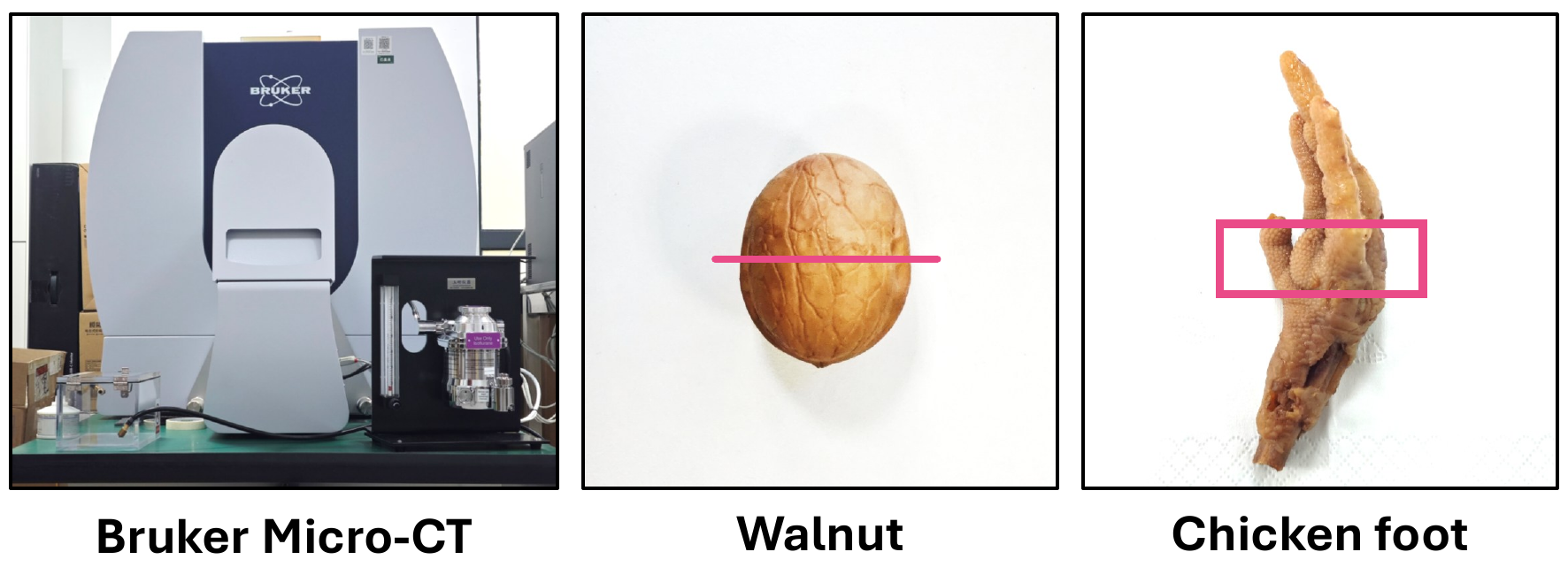}
    \caption{(\textit{Left}) A commercial Bruker SKYSCAN 1276 Micro-CT scanner used to acquire real-world CT projection data. (\textit{Middle} \& \textit{Right}) In-house \textit{Walnut} and \textit{Chicken foot} samples used in our 2D FBCT and 3D CBCT real-world experiments. The pink lines and boxes indicate the regions of interest (ROIs) selected for reconstruction.}
    \label{fig:fig_real_device}
\end{figure}

 For the real-world acquisitions, we use a commercial Bruker SKYSCAN 1276 micro-CT scanner under both 2D FBCT and 3D CBCT protocols to scan two in-house samples, as illustrated in Fig.~\ref{fig:fig_real_device}. Detailed CT acquisition parameters are summarized in Table~\ref{tab:geometry}.
\subsection{Details of Baselines}
\label{sec:sub:baseline}
\par In our experiments, we compare our Riner method with 10 well-known techniques representing the SOTAs for the CT RAR problem. Here, we present the implementation details of these compared methods for improved reproducibility.
\paragraph{FBP/FDK} FBP/FDK~\cite{fbp,fdk} are analytical linear reconstruction algorithms for 2D and 3D CT imaging. In our experiments, we implement them using the Python library \texttt{CIL}~\cite{cil1,cil2} (\url{https://tomographicimaging.github.io/CIL/}).
\paragraph{Norm} Norm, developed by Rivers~\cite{rivers1998Norm}, is a classical CT RAR algorithm. We use the Python library \texttt{Algotom} to implement it.
\paragraph{WaveFFT} WaveFFT~\cite{wavefft} is a well-known filter-based CT RAR algorithm, where raw CT measurements are processed with well-designed filters in the Fourier domain. In our experiment, we use the Python library \texttt{Algotom} to implement it (\url{https://github.com/algotom/algotom}).
\paragraph{Super} Super, proposed by Vo, Nghia T., et al.~\cite{vo2018Super}, represents the SOTA in traditional model-based CT RAR models. This technique comprehensively models various types of ring artifacts and designs specific algorithms to address them. In our experiments, we use its official implementation (\url{https://github.com/algotom/algotom}).
\begin{table*}[t]
    \centering
    \begin{tabular}{lccccc}
    \toprule
      \multirow{2.5}{*}{\textbf{Parameters}} & \multicolumn{3}{c}{\textbf{Simulated Datasets}} & \multicolumn{2}{c}{\textbf{Real-World Datasets}}\\ \cmidrule(r){2-4}\cmidrule(r){5-6}
         & DeepLesion &  LIDC &  AAPM & Walnut & Chicken foot\\ \midrule
         Manufacturer & \texttt{N/A} & \texttt{N/A} & \texttt{N/A} & \texttt{Bruker} & \texttt{Bruker} \\
         Type of geometry & 2D FBCT & 2D FBCT & 3D CBCT & 2D FBCT & 3D CBCT\\
        Image Size & 256$\times$256 & 256$\times$256 & 256$\times$256$\times$100 & 512$\times$512 & 512$\times$512$\times$80\\
        Voxel Size (mm\textsuperscript{2}/mm\textsuperscript{3}) & 1$\times$1 & 1$\times$1 & 1$\times$1$\times$1 & 0.06$\times$0.06 & 0.06$\times$0.06$\times$0.06\\
        View Range ($^\circ$) & [0, 360) & [0, 360) & [0, 360) & [0, 360) & [0, 360)\\
        Source Voltage (kV) & \texttt{N/A} & \texttt{N/A} & \texttt{N/A} & 100& 60 \\
        Source Current (uA) & \texttt{N/A} & \texttt{N/A} & \texttt{N/A} & 200 & 200\\
        Exposure Time (ms) & \texttt{N/A} & \texttt{N/A} & \texttt{N/A} & 276& 250 \\
        Number of Detectors & 500 & 500 & 300$\times$200 & 720 & 120$\times$1008\\
        Detector Spacing (mm) & 2 & 2 & 2 & 0.069 & 0.069\\
        Number of Projection Views & 360 & 360 & 360 & 720 & 720\\
        Distance from Source to Center (mm) & 370 & 370 & 300 & 92.602 & 92.602\\
        Distance from Center to Detector (mm) & 370 & 370 & 300 & 65.946& 65.946 \\
     \bottomrule
    \end{tabular}
    \caption{Detailed parameters of the acquisition protocols for five datasets used in our experiments.}
    \label{tab:geometry}
\end{table*}
\paragraph{HyUNet} Chang et al.\cite{hybridcnn} pioneeringly proposed using convolutional neural networks (CNNs) to learn the unknown mapping from artifact-corrupted images to artifact-free outputs. By constructing a large-scale dataset consisting of extensive pairs of clear and ring-artifact-corrupted images, they trained a five-layer CNN, named HyUNet. In our experiment, we follow the original paper to implement and train the model using the training and validation sets from the DeepLesion dataset\cite{deeplesion}, from scratch with carefully tuned hyperparameters.
\paragraph{DeepRAR} DeepRAR~\cite{trapp2022deeprar} is a supervised DL method designed for the CT RAR problem. It trains a residual UNet~\cite{ronneberger2015u} to map artifact-corrupted CT images to ring artifacts. In our experiments, we reproduce and train it from scratch using the training and validation sets from the DeepLesion dataset~\cite{deeplesion}, with carefully tuned hyperparameters.
\paragraph{NAFNet} NAFNet~\cite{chen2022simple} is a compact and efficient supervised model designed for image restoration tasks such as denoising and deblurring. It is built on the NAFBlock, which simplifies the architecture by removing nonlinear activations and relying on streamlined convolutional and attention-based components. Despite its simplicity, NAFNet achieves competitive performance with significantly reduced computational cost. We use the official implementation (\url{https://github.com/megvii-research/NAFNet}) and train it from scratch on the Deeplesion~\cite{deeplesion} training and validation sets with carefully tuned hyperparameters.
\paragraph{Restormer} Restormer~\cite{zamir2022restormer} is a versatile supervised DL model designed for various low-level vision tasks, such as image denoising and super-resolution. It leverages an efficient ViT~\cite{dosovitskiy2020image} architecture to restore high-resolution images using large-scale paired datasets. Benefiting from the ability of ViT~\cite{dosovitskiy2020image} to capture both local and global image patterns, Restormer~\cite{zamir2022restormer} achieves exceptional reconstruction performance. In our experiments, we utilize its official implementation (\url{https://github.com/swz30/Restormer}) and train it on the training and validation sets from the Deeplesion dataset~\cite{deeplesion} from scratch with carefully tuned hyperparameters. 
\paragraph{MambaIRv2} MambaIRv2~\cite{guo2024mambair,guo2024mambairv2} is a recent SOTA image restoration model that leverages the Mamba-based selective state space mechanism for efficient long-range dependency modeling. Unlike transformer-based methods, MambaIRv2 replaces self-attention with a lightweight state space operator while maintaining spatial attentiveness through gated aggregation and U-shaped designs. This results in improved speed and performance, especially on high-resolution images. We follow the official implementation (\url{https://github.com/csguoh/MambaIR}) and train the model from scratch on the Deeplesion~\cite{deeplesion} training and validation sets using carefully tuned hyperparameters.
% \paragraph{NAF} NAF~\cite{zha2022naf} is an unsupervised INR-based method for 3D CBCT reconstruction. It employs traditional internal models as the CT forward model, allowing reconstruction of 3D volumes from sparse-view projections. In our experiments, we follow its official implementation (\url{https://github.com/Ruyi-Zha/naf_cbct}).
\paragraph{SinoRAR} SinoRAR~\cite{SinoRAR} is an unsupervised INR-based method for 2D CT RAR. Following CoIL~\cite{sun2021coil}, it uses two MLP networks to learn clean CT measurements and strip artifacts. By incorporating handcrafted priors, SinoRAR can recover clean CT measurements, thus enabling CT RAR reconstructions. In our experiments, we reproduce this method following the original paper.
\subsection{Additional Visualization Comparisons}
\label{sec:sub:com}
\par
\begin{figure*}[t]
    \centering
    \includegraphics[width=0.9\linewidth]{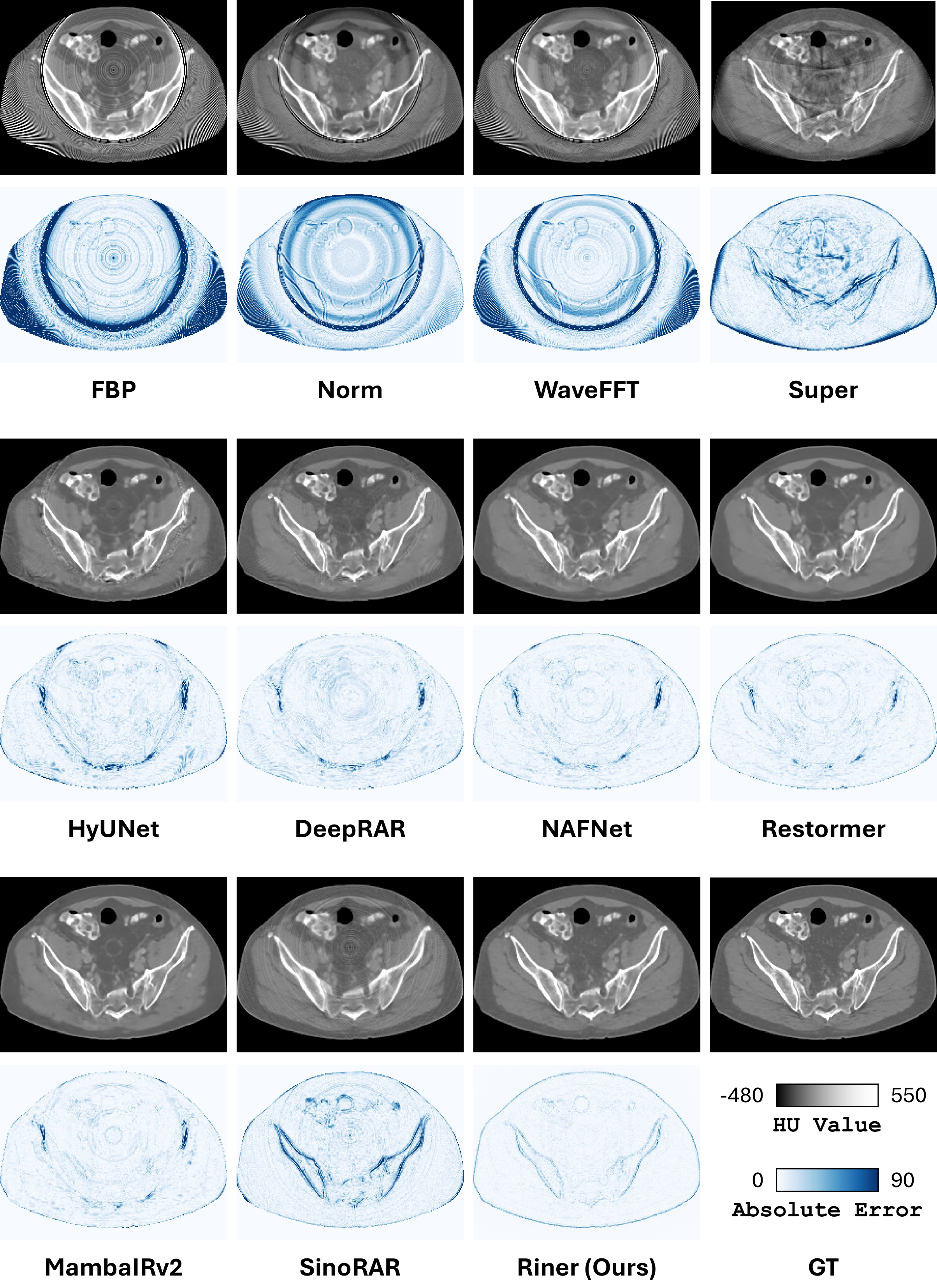}
    \caption{Qualitative results of our Riner and compared methods on a representative sample of the simulated 2D FBCT DeepLesion dataset~\cite{deeplesion}.}
    \label{fig:fig_sub1}
\end{figure*}

\begin{figure*}[t]
    \centering
    \includegraphics[width=0.9\linewidth]{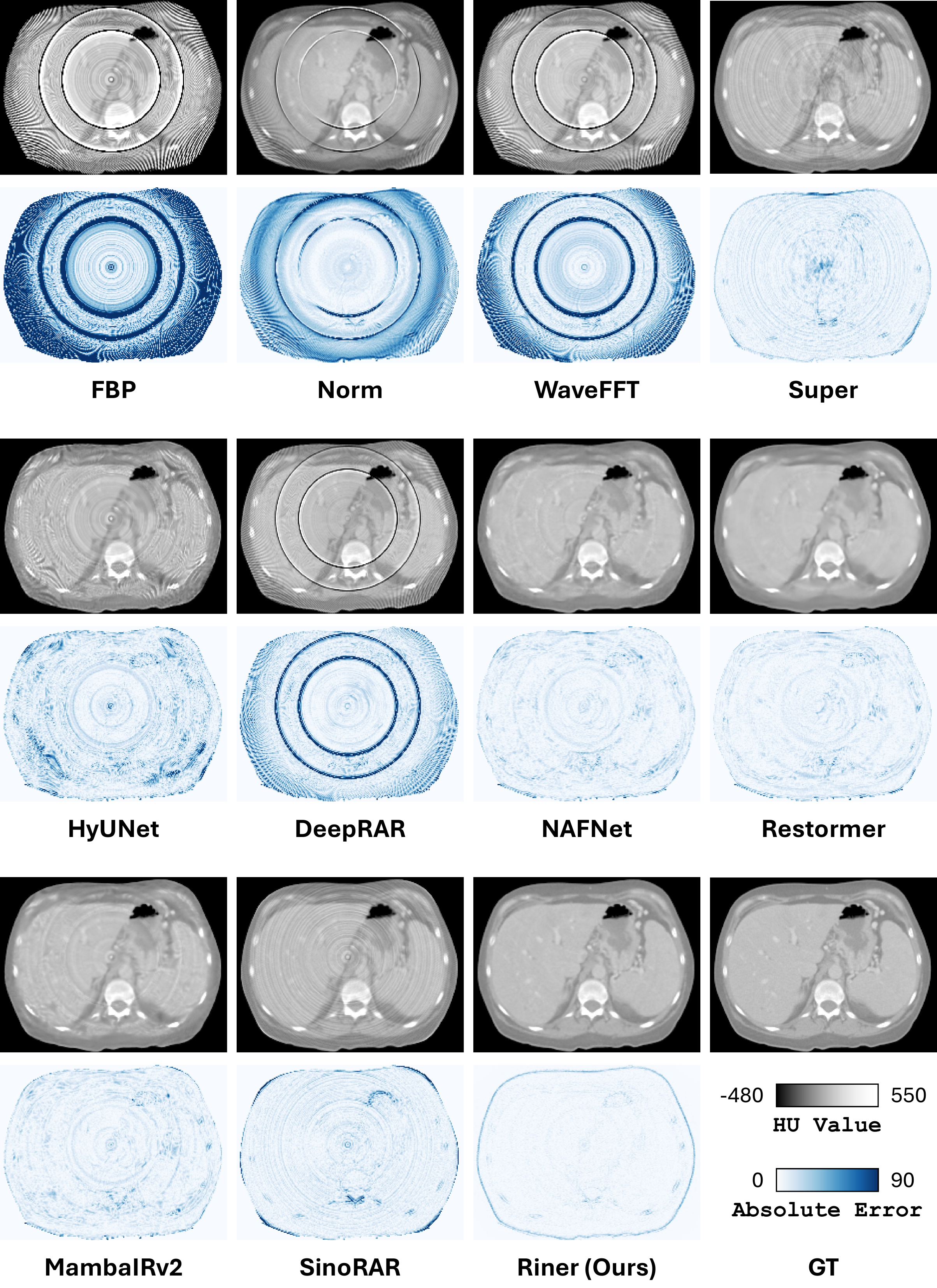}
    \caption{Qualitative results of our Riner and compared methods on a representative sample of the simulated 2D FBCT LIDC dataset~\cite{lidc}.}
    \label{fig:fig_sub2}
\end{figure*}

\begin{figure*}[t]
    \centering
    \includegraphics[width=0.9\linewidth]{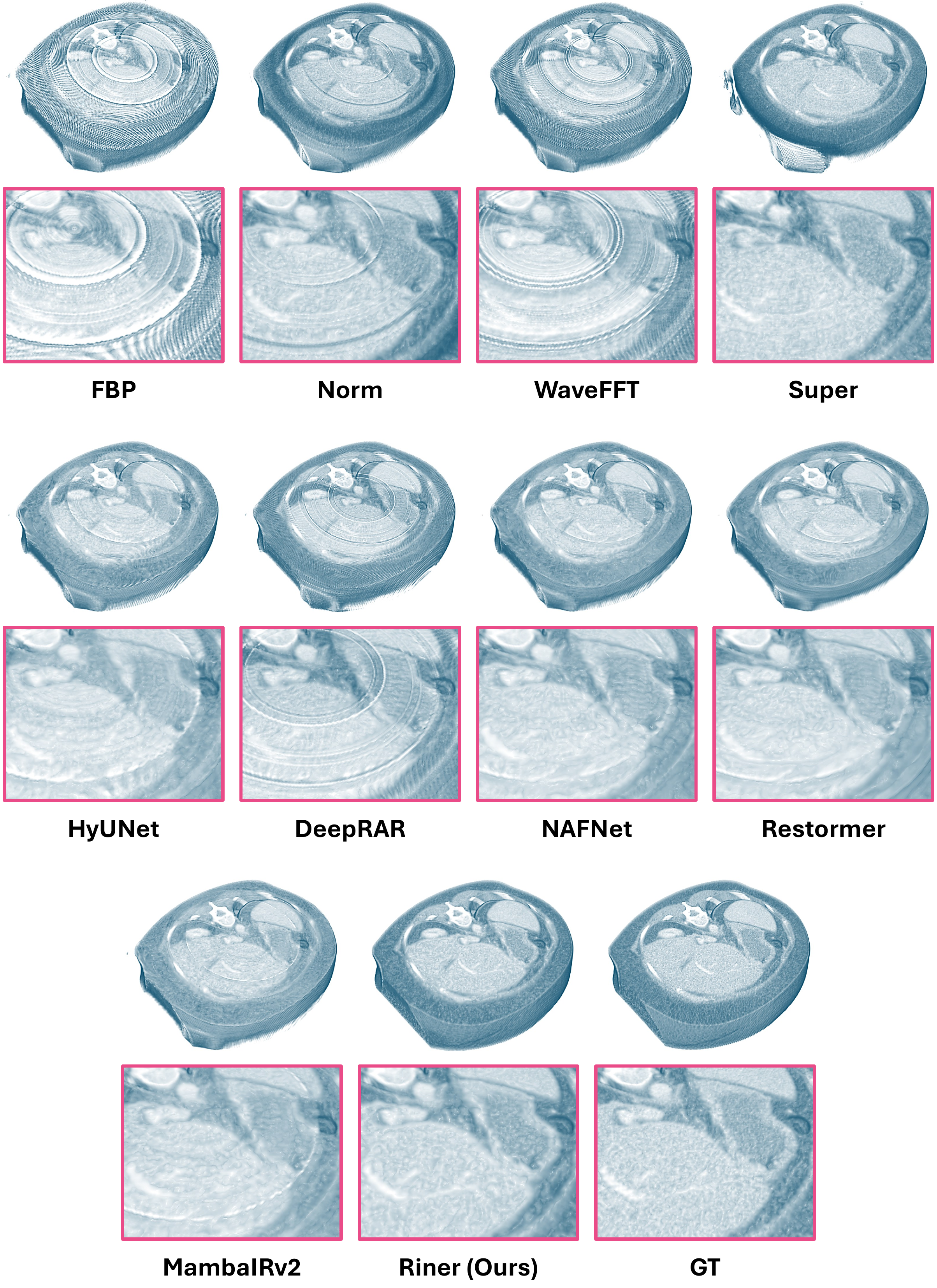}
    \caption{Qualitative results of our Riner and compared methods on a representative sample of the simulated 3D CBCT AAPM dataset~\cite{aapm}.}
    \label{fig:fig_sub3}
\end{figure*}

 Fig.~\ref{fig:fig_sub1}, Fig.~\ref{fig:fig_sub2}, and Fig.~\ref{fig:fig_sub3} show more qualitative comparisons of our Riner and baselines. The proposed unsupervised Riner consistently produces the best RAR reconstructions.

\end{document}